\documentclass{aa}
\usepackage{graphicx}
\usepackage{psfig}
\usepackage{epsfig}
\begin{document}
\def \deg {$^\circ$}
\def\etal{{\it et~al.}}

   \title{The Spiros imaging software for the Integral SPI spectrometer }
  \titlerunning{The Spiros imaging software for SPI}

   \author{G. Skinner
          \inst{1}
          \and
          P. Connell\inst{2}\fnmsep\thanks{Now at : IFC, Universidad de Valencia
E-46100 Burjassot, Valencia, Spain}
          }

   \offprints{G. K. Skinner}

   \institute{C.E.S.R., ,
              9, avenue du Colonel Roche, 31028 Toulouse, France\\
              \email{skinner@cesr.fr}
         \and
             University of Birmingham, Edgbaston, Birmingham, B15 2TT, England\\
             }

   \date{Received .. ; accepted ... }

   \abstract{
A key tool in the package of software available for the analysis
of data from the SPI spectrometer of Integral is the SPIROS system
developed at the University of Birmingham. Although intended
primarily for the analysis of point sources and for the extraction
of spectral information, SPIROS has many additional capabilities.
The software is described with particular emphasis on the most
widely used modes of operation and on the relationship to other
imaging and data reduction techniques.
   \keywords{ Coded masks --
              Imaging --
               Software
               }
   }
   \maketitle
%

\section{Introduction}

   Astronomical images are often of fields consisting of
   (or at least dominated by) a number of sources which are
   essentially points. Their angular size is much smaller than the
   resolution of current instrumentation
    - frequently by a factor more than $10^{10}$.
   At gamma-ray energies the number of such sources
    is usually small.
    In these circumstances the model with a minimum number of
    parameters that is consistent with the data will consist of a list of
    the positions of those sources with their intensities. For an instrument
    with spectroscopic capabilities like INTEGRAL/SPI, a description of the
    variation of those intensities with photon energy, and perhaps with time,
     must be added. In general it is when a minimum number of parameters is
sought  that each of these may be obtained with the highest
    precision\footnote{We note that in a certain sense, this is a `minimum
    information/maximum entropy' interpretation of the
    data - it can be stored or transmitted with a small number of bits
    and hence very high informational entropy as defined by
    Shannon \cite {Shannon}}, so in these circumstances point source searching
    and fitting is the preferred data analysis technique.


\section{Context and environment}
The technique of `Iterative Removal of Sources' (IROS, Hammersley
et al. \cite{hammersly}) has been widely used for coded mask
instruments. A simple image of the field of view is made using a
mapping technique which is optimised for finding a source assuming
that the data can be explained by only that source, plus
background. The mapping gives the approximately location and
intensity of the source, which are then improved by maximising a
measure of the goodness of fit. The residuals of the fit are used
as the input for a further image reconstruction and source search.
The parameters of the two source model are refitted and if the fit
represents a significant improvement on the original one the
process is continued with more and more sources.

SPIROS is a programme which implements this algorithm for the SPI
spectrometer of INTEGRAL, a coded mask imaging instrument with a
detector array comprising 19 high-purity Germanium detectors
giving an angular resolution of about 2.5\deg\ over a field of
view of 16\deg\ (Vedrenne, \etal\ \cite{instrument_paper}). SPIROS
operates within the ISDC software environment (Courvoisier \etal\
\cite {isdc}).


\section{Optimisation criteria and background modelling}

\label{fitting}
The SPI instrument differs from most coded mask
instrument for which IROS has been used because
\begin{enumerate}
  \item The number of detector elements is very limited -- 19 in the
basic mode, though the concept of `pseudo-detectors' can increases
this by up to 104, with double and triple events and pulse shape
discrimination (Vedrenne \etal\ \cite{instrument_paper}).
  \item In consequence multiple telescope pointings
are analysed simultaneously in order to obtain enough information
for a unique solution.
\end{enumerate}

The most widely used coded mask image reconstruction techniques,
based on correlations of the detector plane with a representation
of the form of the mask shadow, usually by FFT, are not
appropriate in such circumstances. The limited number of
measurements means that effects which average out in the
many-pixel limit cannot be ignored. Edge effects and differences
between detectors, in their response for different directions, and
in their noise levels take on a dominant role. However {\it
because} of the limited number of measurements, one can use matrix
techniques that would be impracticable in other cases.

SPIROS operates on data which have been already binned by
(pseudo-) detector, by pointing, and by energy (strictly `pulse
height') bin.  In most modes it treats data from one energy bin at
a time, although a group of input energy bins can be combined to
form one. It also reads files containing attitude information,
integration times, etc, and a file containing one or more
background models.

Considering, then, a particular energy,  for a given set of
assumed source positions, the expected count in detector (or
pseudo-detector) $d$ during pointing $p$ is
\begin{equation}
    {\hat P}_{dp}=t_{dp}\{ \sum_{i}S_i A_{idp}  + B_{dp}\}
    \label{coding_eqn}.
\end{equation}
Here $A_{idp}$ is the effective area of the detector to a source
in the direction of the $i$th source, which is assumed to have
flux $S_i$ within the energy bin considered.
The $A_{idp}$ values are obtained by interpolation of data which
are stored in so-called IRF (Instrument Response Function) files
generated by Monte Carlo simulations.
The effective exposure time, corrected for dead-time and all other
effects, is $t_{dp}$ and the $A_{idp}$ roll together all the
effects of the mask coding and the detector efficiency. For
simplicity, only the photopeak response and a single energy are
considered here.

The possibility that the background is a combination of components
which vary in different ways between pointings and from detector
to detector can be included in this formalism by treating those
components much like sources:
\begin{equation}
    {\hat P}_{dp}=t_{dp}\{ \sum_{i}S_i A_{idp}  + \sum_{i'} F_{i'} B_{i'dp}
    \}.
    \label{coding_eqn_bg}
\end{equation}
Here provision is made for the possibility that the \textit{form}
of the background variation with time may be known but that it
needs to be multiplied by an unknown factor $F_{i'}$, analogous to
a source intensity (see Section \ref{bg_handling}).

The objective is then to find the model (the combination of $S_i$
and, if required, $F_i'$)  which, best explains the observed data
(the count rates ${P}_{dp}$) in the ``Maximum Likelihood" (ML)
sense.

In the general case where the number of counts per bin may be
small, as can be the case when short exposures or narrow energy
bins are being considered, then the relevant
 statistic is (Cash \cite{cash79}) :
\begin{equation}
    C = 2 \sum_{dp} \{ {\hat P}_{dp}- P_{dp}Ln({\hat P}_{dp})\}.
    \label{liklihood_eqn}
\end{equation}
Even in the case of fixed source positions, finding the $S$ and
$F$ which optimises this statistic is a non-linear problem and has
to be  solved iteratively.

If the counts are large, Gaussian statistics can be assumed and
one can minimise the $\chi^2$ statistic
\begin{equation}
    \chi^2 = \sum_{dp} {{(P_{dp}-{\hat P}_{dp})^2}\over{\sigma_{dp}^2}
    }.
    \label{chi2_eqn}
\end{equation}
If the source positions are known (or supposed) and if in addition
one makes the approximation that $\sigma_{dp}^2= {P}_{dp}$,
instead of $\sigma_{dp}^2= {\hat P}_{dp}$  then the problem of
finding the $S_i$ becomes a linear one.

SPIROS can be run either using the general $ML$ approach or using
$\chi^2$ and the assumption that $\sigma_{dp}^2= {P}_{dp}$. The
latter is faster and more efficient in those cases where it can
validly be used. Note that the user must beware of the dangers of
adopting the $\chi^2$ statistic where it is not valid because for
low counts per bin there is a significant probability that
${P}_{dp}=0$. The infinite weights that would result are avoided
by ignoring such data, but this introduces a bias, as does the
incorrect weight given to other bins with low, but non-zero
counts.

\section{Background handling}

\label{bg_handling}

Usually the background has to be treated as an unknown, but some
constraining assumptions are necessary - with $n_p$ pointings and
$n_d$ detectors, it is obviously not possible at the same time to
obtain information about sources in the field of view \emph{and}
make independent estimates of the  background in each of the  $n_d
n_p$ combinations of  detector $d$  and  pointing $p$.

SPIROS reads in one or more ($n_{i'}$, in general) background
models which are sets of $B_{dp}$ values\footnote{strictly
$t_{dp}B_{dp}$ is stored}, generated for a specific data set by an
independent programme (called `spiback'). Sometimes the model
components may be absolute (when they are based on  data preceding
or following an observation, or from energy bands just above and
below  the region of interest, for example). Sometimes they are
simply tracers of a time variation with arbitrary scaling. The
simplest case would be background which is uniform and constant,
corresponding to a single $i'$ component in which all the $B$ have
the same value. Other possibilities are components which depend on
time or from detector to detector in a specific way, or which
follow tracers of expected background contributions (for example
the veto shield count rate is a measure of the particle flux, the
rate of out of range events in the germanium detectors is a
measure of high energy particles\ldots).

In its existing form SPIROS has provides for the following
possibilities :
\begin{enumerate}
\item{ The backround estimates may be taken to be absolute
estimates, not requiring the fitting of a factor $F$} \item{They
may be taken to reflect the time (pointing to pointing) variation
of the background but with a potentially different scaling factor
$F$ per detector, to be found by fitting within SPIROS.} \item{It
may be assumed that relative, detector-detector, backgrounds have
been provided, as well as the correct time ($p$) dependence, so
that SPIROS need only fit a single normalising  $F$.} \item{In a
slight generalisation of (3), one factor may be fitted for all
single detectors but a different factor for pseudo-detectors
corresponding to double events in particular pairs of detectors
(or triples, etc.)}
\end{enumerate}

The most general case (2) is usually adopted. Further options may
be added in the future, such as assuming that the relative
backgrounds in different detectors are known but with a different
unknown scaling factor to be fitted per pointing or on some other
timescale.

\section{Mapping; finding and positioning sources}

A basic mapping operation consists of considering each pixel in
the image successively, placing a test source at that position and
establishing the intensity that it would have in order to best
match the observed data, along with the uncertainty in that
intensity. Sources found in previous iterations, or read in from a
catalogue, are either subtracted out from the data ($\chi^2$) or
taken into account in the analysis ($ML$). For source searching
the intensities and uncertainties evaluated on a comparatively
coarse grid (\textit{e.g.} 0.5\deg ) can be used. A smoothed
linear interpolation is then used to fill in estimates on a finer
pitch.

The source selected for potential addition to the list of sources
is that which has the highest value of intensity/uncertainty.

Before accepting a new source as real, a simultaneous optimisation
of its position \textit{and reoptimisation of all the other
sources that do not have good catalogued positions} is performed
(section \ref{relationship} below). The procedure used is an
iterative one with a descent along the line of maximum slope.

\section{Relationship of IROS to other methods}
\label{relationship} Numerous  deconvolution techniques have been
proposed and used for image reconstruction in astronomy. Examples
are Maximum Entropy, and Richardson-Lucy; see Starck \& Pantin
(\cite{decon-review}) for a recent review. Images obtained by such
methods can be considered as vectors in an \textit{N-}dimensional
space, where the component of the vector along each of the axes
represents the intensity in one of the \textit{N} pixels. In the
presence of noise, there will always be a certain volume in this
space, inside which the points correspond to images that are
consistent with the data. Different methods apply constraints
which are not the same and so do not lead to the same choice of a
point in hyperspace among all of the possible ones.

The IROS algorithm searches the space for a solution consistent
with the data according to the following rules:
\begin{description}
  \item[\textit{Mapping :}] Starting from point in this space which corresponds
  to the current solution, test each axis (pixel number)
  to find the point along that axis with the lowest $\chi^2$.
  \item[\textit{Source finding :}] Identify which pixel gave the maximum
  descent in the sense corresponding to a positive source flux.
  Step along the axis corresponding to that pixel to the lowest
  point; all other components of the vector are left at their original
values.
  \item[\textit{Fitting :}] Intensity fitting corresponds to optimising in a subspace
  which has just those dimensions
  corresponding to the sources already found.
  The position fitting is more complex. An optimisation is made in a
  separate space with $3S$ dimensions, corresponding to the $2$ position coordinate
  and the intensity of each of $S$ sources. The $S$ directions corresponding
  to the optimised positions of the sources are then added to the
  original space to make an $N+S$ dimensional one.

\end{description}

Formally, there is no guarantee that this algorithm leads to a
minimal description of the data in terms of point sources.
Pragmatically, in simulations and in observations of involving
known sources, it is found to be effective.

The procedure is very similar to the CLEAN method used in radio
astronomy (H\"ogbom, \cite{hogbom}; Schwarz, \cite{schwarz}),
except that in CLEAN (i) position optimisation is not normally
performed, (ii) the point in the map with the highest
\textit{absolute} value is chosen, so iterations can add
\textit{negative} components to the image, (iii) only a fraction
of the intensity of a source is subtracted. The resulting image
will usually have a relatively large number of non-zero pixels,
whereas with IROS, there is just one per source. CLEAN can be
considered as an exploration of the multi-dimensional space
considered above, with no positivity constraint and without the
fitting stage.

The IROS algorithm differs from many image reconstruction
techniques in that the possible source positions are not
restricted to a fixed grid of pixel positions.

\section{Spectral extraction}

In a different mode, SPIROS is used for the extraction of spectral
information. This requires an input catalogue, which may simply be
a list of known sources or which may be the result of a previous
run of SPIROS in imaging mode. In each energy bin in turn, the
combination of source intensities (and, if required, of background
parameters) which best matches the data is found. The method is as
described in section \ref{fitting}.

In this way one obtains for each source a spectrum 
analogous to a `Pulse Height' spectrum in that off-diagonal terms
in the energy response matrix are not taken into account. However
other aspects of the instrument response (detector photopeak
efficiency variation with energy, for example) have been corrected
for. For sources with conventional continuum spectra, this measure
is already a very good estimate of the input spectrum. However for
definitive results, a programme such as XSPEC (Arnaud
\cite{xspecref}) needs to be used to take into account the
off-diagonal terms in the response of the combination, SPI+SPIROS.
XSPEC-compatible response matrices for this step have been derived
using
Monte Carlo simulations of observations of monoenergetic sources
at 100 different energies.

\section{Other modes}

Optionally, the sources may be treated as having a finite extent.
For example it is possible to treat each source as a Gaussian
function and find the width which best matches the data.

Extracting a light curve (intensity as a function of time) for
each of the sources in an input catalogue is directly analogous to
the extraction of pulse height spectra.

Although SPIROS is intended for fitting of point sources, it does
have a mode in which one solves simultaneously for the intensities
of the fluxes in each pixel of an image, allowing map of diffuse
emission to be generated. Such inverse problems are notoriously
unstable if the number of pixels is high and the coding is not
ideal. Thus instead of a simply multiplying by the inverse of the
coding matrix, it is modified by the addition of a diagonal Wiener
term or some other smoothing constraint matrix, giving a
stabilising effect by allowing the diagonal terms to dominate.

\section{Examples of results obtained with SPIROS and Conclusions}
   \begin{figure}
\vspace{0cm} %
\hspace{0cm}\psfig{figure=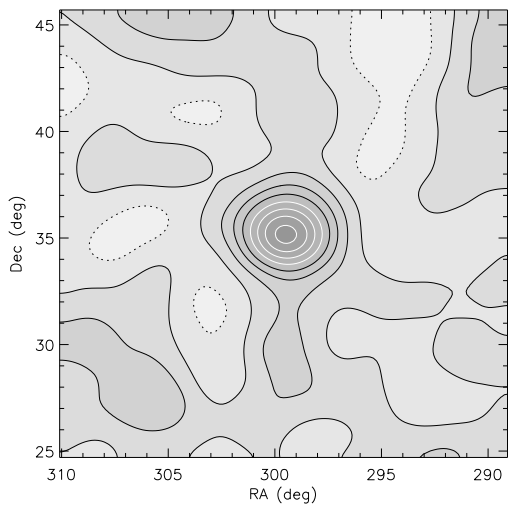,width=10cm,angle=0}
\vspace{0cm}
      \caption[]{An image of Cygnus X-1 obtained by using double and triple events.
      The energy range is 100--700 keV, but most of the events are $>$300 keV.
      Contours are at intervals  of 4$\sigma$; dashed lines are negative.
      The peak corresponds to 29$\sigma$.}
         \label{cygx1-multiples}
   \end{figure}
%

   \begin{figure}
\vspace{0cm} %
\hspace{0cm}\psfig{figure=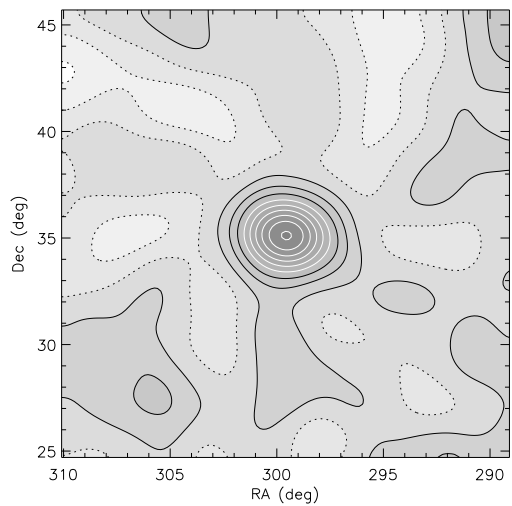,width=10cm,angle=0}
\vspace{0cm}
      \caption[]{As Figure \ref{cygx1-multiples} for single events. Note
      that the intensity scale is not the same: contour
      intervals are 14$\sigma$ and the peak is 134$\sigma$.
      }
         \label{cygx1-singles}
   \end{figure}
%

%

Results obtained with Spiros can be seen in other articles in this
issue and elsewhere - see, for example, Bouchet \etal~
(\cite{bouchet}).

As an example of a slightly non-standard use of Spiros, Figure
\ref{cygx1-multiples} shows an image obtained using \emph{only}
events interacting in 2 or 3 detectors. For such events there is
no information about which of the 2 or 3 detectors corresponds to
the first interaction.  The source is well identified and located,
though with significance lower than in the corresponding single
event image (Figure \ref{cygx1-singles}).

The form in which the IROS  algorithm is implemented within SPIROS
handles for the first time  the situation where no simplifying
assumptions can made about position independence of the recorded
mask shadow. By using the instrument response characterised by a
generalised matrix and an object-oriented programme structure,
additional sophistication can be introduced as 
necessary and as the knowledge of the instrument improves.

\begin{acknowledgements}
      The development of SPIROS was supported by a grant from
      PPARC. The work has benefited from inputs from many
      members of the Integral SPI Data Analysis Group (ISDAG)
      and the Integral Science Data Centre (ISDC).
\end{acknowledgements}


\begin{thebibliography}{}

\bibitem[1996]{xspecref} Arnaud, K. A. 1996,   %
      ASP Conf. Series, 101, 17

  \bibitem[2003]{bouchet} Bouchet, L. , Jourdain, E., Roques, J.-P., \etal , 2003,
      This volume

  \bibitem[1979]{cash79} Cash, W.  1979,
      Ap.J, 228, 939

 \bibitem[2003]{isdc} Couvouisier, T., Walter, R., Beckmann, V.,  \etal , 2003,
      This volume.

   \bibitem[1984]{hammersly} Hammersly, A. , Ponman, T., \& Skinner, G. 1992,
      NIMS, A311, 585

   \bibitem[1974]{hogbom} H\"ogbom, J. A. 1974,
      A\&AS, 15, 417

  \bibitem[1978]{schwarz} Schwarz, U. J. 1978,
      A\&A, 65, 345

 \bibitem[1966]{Shannon} Shannon, H. 1948,
      Bell Syst. Tech. J. 27, 379

 \bibitem[2002]{decon-review} Starck, J. L., Pantin, E., \& Murtagh, F. 2002,   %
      PASP, 114, 1051


 \bibitem[2003]{spiskymax} Strong, A. W. 2003,   %
      This volume.

 \bibitem[2003]{resp_gen} Sturner, S., Shrader, C., Weidenspointner, G., \etal , 2003,
        This volume.


  \bibitem[2003]{instrument_paper} Vedrenne, G., Roques, J.-P., Sch\"onfelder, V.,  \etal , 2003,
      This volume.

\end{thebibliography}
\end{document}